\documentclass[letter]{aa}

\usepackage{natbib}
\usepackage{graphicx}
\usepackage{txfonts}
\usepackage[bookmarks=false, colorlinks=true, citecolor=blue, linkcolor=blue]{hyperref}
\usepackage{subfig}

\def\micron{\hbox{\,$\mu$m}}
\newcommand{\Lsun}{\hbox{$L_{\rm \odot}$}}
\newcommand{\LIR}{\hbox{$L_{\rm IR}$}}
\newcommand{\degree}{\ensuremath{^\circ}}
\newcommand{\waterline}{\hbox{$4_{23}-3_{30}$}}

\defcitealias{Pereira2016b}{PS16}
\defcitealias{GonzalezAlfonso2014}{GA14}

\bibpunct{(}{)}{;}{a}{}{,}

\DeclareCaptionListOfFormat{subreformat}{#1#2}
\titlerunning{Water \waterline\ 448\,GHz transition in ESO~320-G030}
\authorrunning{Pereira-Santaella et al.}

\begin{document}

\title{First detection of the 448\,GHz H$_2$O transition in space}

\author{M.~Pereira-Santaella\inst{\ref{inst1}} \and E.~Gonz\'alez-Alfonso\inst{\ref{inst2}} \and A.~Usero\inst{\ref{inst3}} \and S.~Garc\'ia-Burillo\inst{\ref{inst3}} \and J.~Mart\'in-Pintado\inst{\ref{inst4}} \and L.~Colina\inst{\ref{inst4}} \and A.~Alonso-Herrero\inst{\ref{inst4}} \and S.~Arribas\inst{\ref{inst4}} \and S.~Cazzoli\inst{\ref{inst5}}  \and F.~Rico\inst{\ref{inst4}} \and D.~Rigopoulou\inst{\ref{inst1}} \and T.~Storchi Bergmann\inst{\ref{inst6}} 
}

\institute{Department of Physics, University of Oxford, Keble Road, Oxford OX1 3RH, UK\\ \email{miguel.pereira@physics.ox.ac.uk}\label{inst1}
\and 
Universidad de Alcal\'a, Departamento de F\'isica y Matem\'aticas, Campus Universitario, 28871 Alcal\'a de Henares, Madrid, Spain\label{inst2}
\and
Observatorio Astron\'omico Nacional (OAN-IGN)-Observatorio de Madrid, Alfonso XII, 3, 28014, Madrid, Spain\label{inst3}
\and
Centro de Astrobiolog\'ia (CSIC/INTA), Ctra de Torrej\'on a Ajalvir, km 4, 28850, Torrej\'on de Ardoz, Madrid, Spain\label{inst4}
\and
Instituto de Astrof\'isica de Andaluc\'ia, CSIC, Glorieta de la Astronom\'ia, s/n, E-18008 Granada, Spain\label{inst5}
\and
Universidade Federal do Rio Grande do Sul, Instituto de F\'isica, CP 15051, Porto Alegre 91501-970, RS, Brazil\label{inst6}
}

\abstract{
We present the first detection of the ortho-H$_2$O \waterline\ transition at 448\,GHz in space. We observed this transition in the local ($z=0.010$) luminous infrared (IR) galaxy ESO~320-G030 (IRAS~F11506-3851) using the Atacama Large Millimeter/submillimeter Array (ALMA). The water \waterline\ emission, which originates in the highly obscured nucleus of this galaxy, is spatially resolved over a region of $\sim$65\,pc in diameter and shows a regular rotation pattern compatible with the global molecular and ionized gas kinematics.
The line profile is symmetric and well fitted by a Gaussian with an integrated flux of 37.0 $\pm$ 0.7\,Jy\,km\,s$^{-1}$. 
Models predict this water transition as a potential collisionally excited
maser transition. On the contrary, in this galaxy, we find that the
\waterline\ emission is primarily excited by the intense far-IR radiation
field present in its nucleus. According to our modeling, this transition is a
probe of deeply buried galaxy nuclei thanks to the high dust optical depths ($\tau_{100\mu m}>1$, $N_{\rm H}>10^{24}$\,cm$^{-2}$) required to efficiently excite it.}
\keywords{galaxies: ISM -- galaxies: nuclei -- infrared: galaxies -- ISM: molecules}

\maketitle
 
\section{Introduction}\label{sec:intro}

Water is a molecule of astrophysical interest because it not only plays a central role in the Oxygen chemistry of the interstellar medium (e.g., \citealt{Hollenbach2009, vanDishoeck2013}) but it is also one of main coolants of shocked gas (e.g., \citealt{Flower2010}). In addition, thanks to its energy level structure, water couples very well to the far-infrared (far-IR) radiation field providing an effective probe of the far-IR continuum in the warm compact regions found in active galactic nuclei (AGN) and young star-forming regions (e.g., \citealt{GonzalezAlfonso2014}, hereafter \citetalias{GonzalezAlfonso2014}).

\begin{figure}
\centering
\vspace{5mm}
\includegraphics[width=0.42\textwidth]{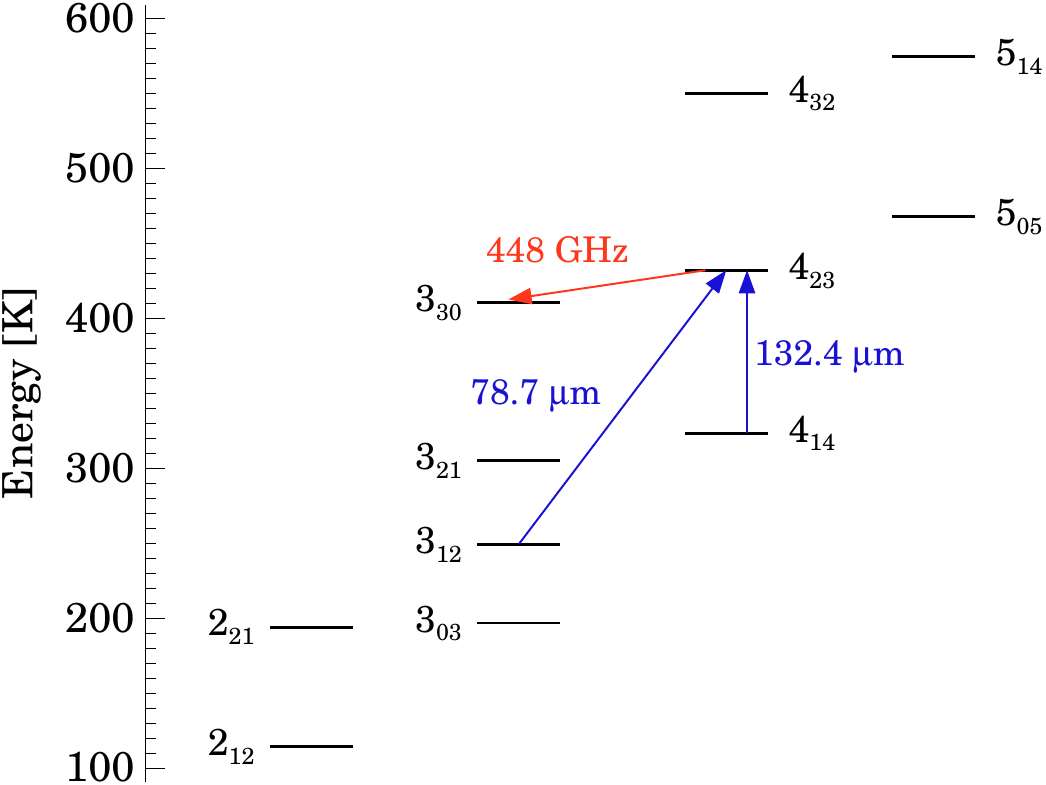}
\caption{Partial energy level diagram of ortho-H$_2$O. The \waterline\ 448\,GHz transition is indicated in red. The 78.7 and 132.4\micron\ transitions, which populate the 4$_{23}$ level radiatively through the absorption of far-IR photons (see Section~\ref{s:model}), are marked in blue. \label{fig:levels}}
\end{figure}

Water excitation models have long predicted the maser nature of the \waterline\ transition pumped by collisions when the kinetic temperature is $T_{\rm kin}\sim1000\,{\rm K}$ and the hydrogen density $n_{\rm H_2}\sim10^5$\,cm$^{-3}$ 
(e.g., \citealt{Deguchi1977,Cooke1985,Neufeld1991,Yates1997,Daniel2013,Gray2016}). 
This transition can also be excited by radiative pumping through the absorption of far-IR photons (see Section~\ref{s:model} and Figure~\ref{fig:levels}). Therefore, the determination of the dominant excitation mechanism, which might vary from source to source, is required to properly interpret the \waterline\ emission as a tracer of dense hot molecular gas or as a tracer of intense IR radiation fields in compact regions.

In this letter, we present the first detection of the ortho-H$_2$O \waterline\ 448.001\,GHz transition in space\footnote{\citet{Persson2007} reported a tentative detection of the water isotopologue H$_2^{18}$O
\waterline\ transition at 489.054\,GHz in Orion, although it is blended with a much stronger methanol transition.}. No previous detections of this transition in Galactic objects have been reported, probably because of the high atmospheric opacity due to the terrestrial water vapor.
Only recently, thanks to the sensitivity of the Atacama Large Millimeter/submillimeter Array (ALMA), it became possible to observe this transition in nearby galaxies red-shifted into more accessible frequencies.

We observed the H$_2$O \waterline\ transition in ESO~320-G030 (IRAS~F11506-3851; $D = 48$\,Mpc; 235 pc\,arcsec$^{-1}$). This object is an isolated spiral galaxy with a regular velocity field \citep{Bellocchi2016} and an IR luminosity (log \LIR\slash\Lsun = 11.3) in the lower end of the luminous IR galaxy (LIRGs) range (11 $<$ log \LIR\slash\Lsun $<$ 12). It is a starburst object with no evidence of an AGN based on X-ray and mid-IR diagnostics (\citealt{Pereira2010, Pereira2011}) hosting an extremely obscured nucleus ($A_{\rm V}\sim40$\,mag)
and a massive outflow powered by the presumed nuclear starburst detected in the ionized, neutral atomic and molecular phases (\citealt{Arribas2014}; \citealt{Cazzoli2014, Cazzoli2016}; \citealt{Pereira2016b}, hereafter \citetalias{Pereira2016b}). In addition, a molecular gas inflow is suggested by the inverse P-Cygni profile observed in the far-IR OH absorptions \citep{GonzalezAlfonso2017}. It is an OH megamaser source \citep{Norris1986}, but no 22\,GHz H$_2$O maser emission has been detected \citep{Wiggins2016}. This is consistent with the starburst activity of the nucleus of ESO~320-G030 (see \citealt{Lo2005}).

\section{ALMA data reduction}\label{s:data}

We obtained band 8 ALMA observations of ESO~320-G030 on 2016 November 16 using 42 antennas of the 12-m array as part of the project \#2016.1.00263.S. The total on-source integration time was 10.5\,min. The baselines ranged from 15\,m to 920\,m that correspond to a maximum recoverable scale of $\sim$2\arcsec\ based on the ALMA Cycle\,4 Technical Handbook equations. A three pointing pattern was used to obtain a mosaic with uniform sensitivity over a $\sim$8\arcsec$\times$8\arcsec\ field of view.

In this letter, we only use data from a spectral window centered at 443.0\,GHz (1.875\,GHz\slash 1270\,km\,s$^{-1}$ bandwidth and 1.95\,MHz\slash 1.3\,km\,s$^{-1}$ channels) were the redshifted H$_2$O \waterline\ 448.001\,GHz transition is detected. The remaining ALMA data will be analyzed in a future paper (Pereira-Santaella et al., in prep.)
The data were reduced and calibrated using the ALMA reduction software CASA (v4.7.0; \hbox{\citealt{McMullin2007}}). 
For the flux calibration we used J1229+0203 (3C~273) assuming a flux density of 2.815\,Jy at 449.6\,GHz and a spectral index $\alpha=-0.78$ ($f_\nu\propto\nu^\alpha$). The final data-cube has 300$\times$300 pixels of 0\farcs05 and 31.2\,MHz (20\,km\slash s) channels. For the cleaning, we used the Briggs weighting with $R=0.5$ \citep{Briggs1995PhDT} which provides a beam with a full-width half-maximum (FWHM) of 0\farcs26$\times$0\farcs24 ($\sim$60\,pc) and a position angle (PA) of 58\degree. The 1$\sigma$ sensitivity is $\sim$4.8\,mJy\,beam$^{-1}$ per channel. We corrected the data-cube for the primary beam pattern of the mosaic.

\section{Data analysis}\label{s:analysis}

\begin{figure}
\centering
\includegraphics[width=0.44\textwidth]{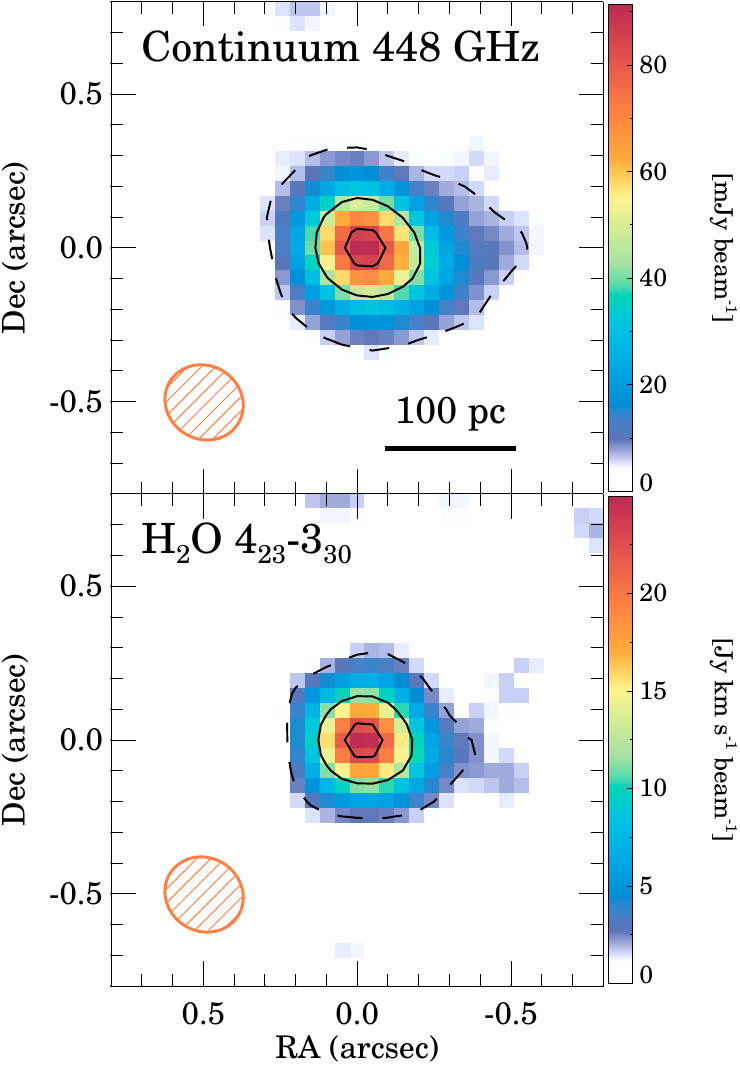}
\caption{Map of the 448\,GHz (rest frequency) continuum (top panel) and zeroth moment of the H$_2$O \waterline\ emission (bottom panel) of ESO~320-G030. The dashed line contour marks the 3$\sigma$ level (7\,mJy\,beam$^{-1}$ and 2.5\,Jy\,km\,s$^{-1}$\,beam$^{-1}$, respectively). The solid contour lines indicate the peak$\times$(0.5, 0.9) levels. The red hatched ellipses indicate the beam size (0\farcs26$\times$0\farcs24, PA$=58\degree$). The coordinates are relative to 11 53 11.7192 +39 07 49.105 (J2000).
\label{fig:alma_maps}}
\end{figure}

\begin{figure}
\centering
\includegraphics[width=0.38\textwidth]{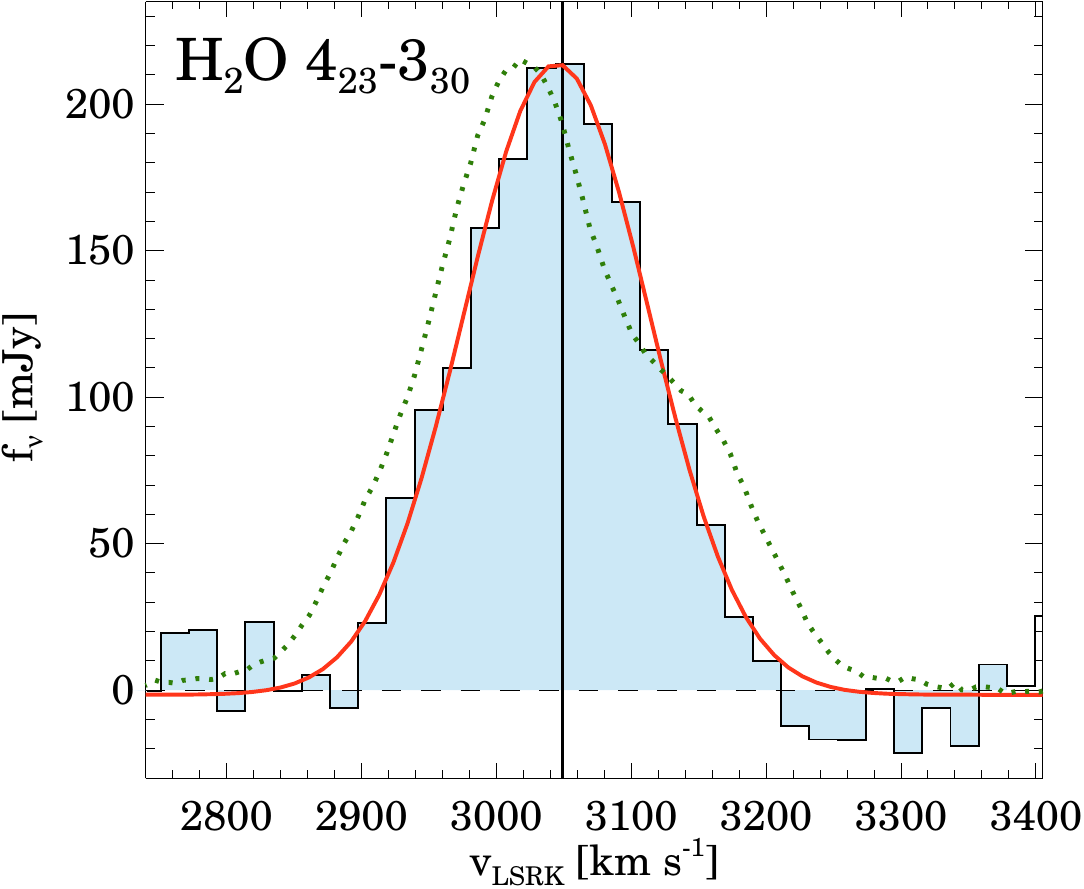}
\caption{Continuum subtracted profile of the H$_2$O \waterline\ 448\,GHz emission in ESO~320-G030 extracted using a circular aperture with $d=0\farcs8$ centered at the nucleus (see Figure~\ref{fig:alma_maps}). 
The dotted green line is the normalized CO(2--1) profile extracted from the same region.
The black vertical line indicates the systemic velocity derived from the CO(2--1) global kinematic model \citepalias{Pereira2016b}.
The red solid line is the best Gaussian fit to the water profile (see Section~\ref{s:analysis}). 
\label{fig:h2o_profile}}
\end{figure}

We detect continuum and line emission only in the central $\sim$200\,pc ($\sim$1\arcsec). This is consistent with the extent of the 233\,GHz (1.3\,mm) continuum emission in this object (see \citetalias{Pereira2016b}).
We estimated the continuum level in each pixel from the median flux density in the line-free channels of the spectral window. The resulting continuum map is shown in Figure~\ref{fig:alma_maps}. 
The measured total continuum emission in the central 200\,pc is 183 $\pm$ 4\,mJy.

From the continuum subtracted data cube, we extracted the nuclear spectrum using a $d=0\farcs8$ aperture (Figure~\ref{fig:h2o_profile}). A line is detected at 443451 $\pm$ 2\,MHz. This corresponds to a rest frame frequency of 448007 $\pm$ 4\,MHz (using the systemic velocity $v_{\rm radio} = 3049\pm2$\,km\,s$^{-1}$, derived from CO(2--1); see \citetalias{Pereira2016b}) which agrees with the frequency expected for the ortho-H$_2$O \waterline\ transition (448001\,MHz; \citealt{Pickett1998}). This line identification is also supported by the detection of strong far-IR and sub-mm water transitions in the \textit{Herschel} observations of this object (see Section~\ref{s:model}). 
We also detect a weaker emission line (3.7$\pm$0.6\,Jy\,km\,s$^{-1}$) which we tentatively identify as two CH$_2$NH transitions at $\sim$446.8\,GHz ($E_{\rm up}$=96 and 117\,K; \citealt{Pickett1998}).
Another two CH$_2$NH transitions at 447.9 and 448.1\,GHz might contribute to the \waterline\ flux. But they have higher $E_{\rm up}$, $\geq 280$\,K, so their contributions are likely negligible.

We fitted a Gaussian to the H$_2$O \waterline\ profile and the result is shown in Figure~\ref{fig:h2o_profile}. We obtained a total flux of 37.0 $\pm$ 0.7\,Jy\,km\,s$^{-1}$, a velocity of 3045 $\pm$ 1\,km\,s$^{-1}$, and a FWHM of 161 $\pm$ 2\,km\,s$^{-1}$. The \waterline\ profile is symmetric and it is centered at the systemic velocity. By contrast, the nuclear CO(2--1) profile has a higher FWHM and presents a more complex asymmetric profile (see Figure\,\ref{fig:h2o_profile} and figure\,6 of \citetalias{Pereira2016b}).

From the 448\,GHz continuum and the zeroth moment water emission maps (Figure~\ref{fig:alma_maps}), we measured the sizes of the emitting regions by fitting a 2D Gaussian. Both the continuum and the water emission are spatially resolved in the ALMA observations with the continuum being more extended. The continuum size (FWHM) is 0\farcs38$\times$0\farcs32, which, deconvolved by the beam size, corresponds to 60\,pc$\times$50\,pc at the distance of ESO~320-G030. The size of the water emission is 0\farcs30$\pm$0\farcs02, which is equivalent to a deconvolved FWHM of 40$\pm$3\,pc.
For a uniform-brightness disk, the equivalent radius is $0.8\times{\rm FWHM}$ \citep{Sakamoto2008}, i.e., $R\sim45$ and $30-35$\,pc for the 448 GHz continuum and the H$_2$O line, respectively.

\begin{figure}
\centering
\vspace{5mm}
\includegraphics[width=0.43\textwidth]{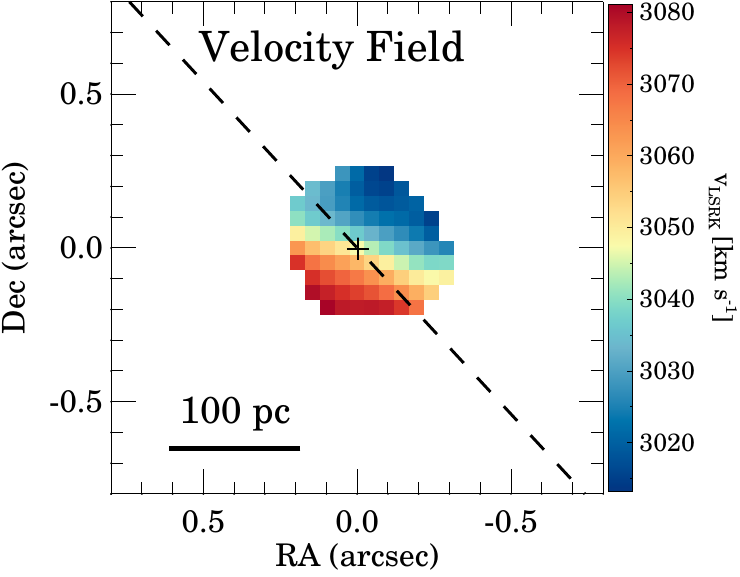}
\caption{Velocity field of the H$_2$O \waterline\ emission. The black cross marks the position of the water emission peak (see Figure~\ref{fig:alma_maps}). The dashed line is the minor kinematic axis derived from the kinematic analysis of the CO(2--1) emission (see \citetalias{Pereira2016b}).
\label{fig:h2o_kinematics}}
\end{figure}

We also determined the spatially resolved kinematics of the water emission by fitting a Gaussian profile pixel by pixel. The velocity field 
of the water line is shown in Figure~\ref{fig:h2o_kinematics} for the pixels where the line is detected at $>$3$\sigma$. It shows a clear rotating pattern whose kinematic axes are approximately aligned with the large-scale kinematic axes derived from both the CO(2--1) and H$\alpha$ emissions (\citetalias{Pereira2016b}; \citealt{Bellocchi2013}). The slight angular deviation, $\sim$25\degree, is similar to that observed in the nuclear CO(2--1) kinematics and it might be related to the secondary stellar bar and the elongated molecular structure associated with this bar \citepalias{Pereira2016b}. The FWHM line widths ranges from $\sim$100--170\,km\,s$^{-1}$ with the maximum value close to the water emission peak.

Based on the measured continuum fluxes at 448\,GHz and 244\,GHz (\citetalias{Pereira2016b}), and on the emitting region size, we estimated the dust temperature and optical depth. First, we subtracted the free-free contribution at these frequencies ($\sim$7\,mJy; \citetalias{Pereira2016b}). Then, we solved the gray-body equation assuming $1.5<\beta<1.85$ and using a Monte Carlo bootstrapping method to estimate the confidence intervals. We find that $T_{\rm dust}=25-80$\,K and $\tau_{\rm 448\,GHz}=0.2-1.3$. These values may be significantly higher in the more compact region sampled by the H$_2$O 448\,GHz emission.

\begin{figure}
\centering
\vspace{5mm}
\subfloat{\label{fig:model_a}}
\subfloat{\label{fig:model_b}}
\subfloat{\label{fig:model_c}}
\includegraphics[width=0.43\textwidth]{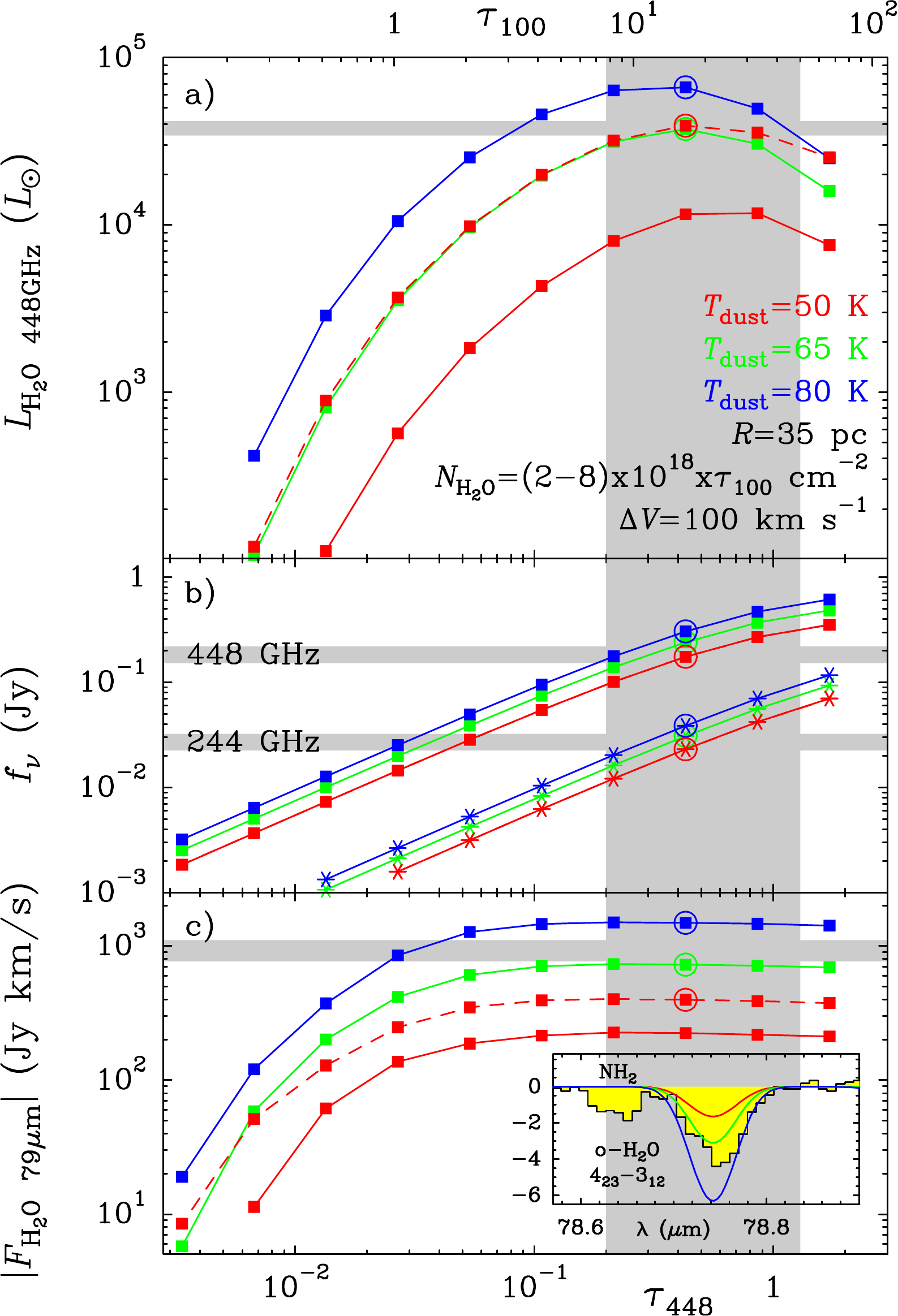}
\caption{a) Model predictions showing the luminosity of the 
H$_2$O 448\,GHz line as a function of the continuum
optical depth at 448\,GHz ($\tau_{\rm 448}$, lower axis) and at 100\,$\mu$m
($\tau_{\rm 100}$, upper axis), for uniform $T_{\mathrm{dust}}=50$, 65, and
80\,K. The models assume spherical symmetry with a radius $R=35$ pc.
The assumed H$_2$O abundance is $X(\mathrm{H_2O})=1.5\times10^{-6}$ (solid
lines) and $X(\mathrm{H_2O})=6\times10^{-6}$ (dashed red line). 
The shadowed regions mark the favored ranges inferred from the ESO~320-G030
observations. 
b) Comparison between the predicted continua at 448\,GHz (squares) and
244\,GHz (starred symbols) and the observed values (after subtracting the
free-free emission; horizontal stripes). 
c) Comparison between the predicted absorbing flux of the pumping H$_2$O
$4_{23}-3_{12}$ line at 79\,$\mu$m and the observed value ($-920$
Jy\,km\,s$^{-1}$ within $\pm150$ km\,s$^{-1}$; horizontal stripe).
The insert compares the observed H$_2$O $4_{23}-3_{12}$ absorption at
78.7\,$\mu$m with the predictions of the three models encircled in the three
panels. 
The width of the horizontal stripes assume uncertainties of $\pm10$\% 
for $L_{\rm{H_2O\,448}}$, and $\pm20$\% for the continuum flux densities and 
for the flux of the H$_2$O 79\,$\mu$m line.
\label{fig:model}}
\end{figure}

\section{Modeling the H$_2$O 448\,GHz emission}\label{s:model}

Figure~\protect\subref{fig:model_a} shows the model predictions 
for the H$_2$O\,448\,GHz luminosity 
as a function of the continuum optical depth for different
dust temperatures ($T_{\mathrm{dust}}=50$, 65, and 80\,K). 
The models, based on those reported in \citetalias{GonzalezAlfonso2014},
use the observed size ($R=35$\,pc) and assume a H$_2$O column density of 
$N_{\mathrm{H_2O}}=2\times10^{18}\times\tau_{100}$ (solid lines) and
$8\times10^{18}\times\tau_{100}$ (dashed red line). These values
correspond to H$_2$O abundances relative to H nuclei of 
$X_{\mathrm{H_2O}}=1.5\times10^{-6}$ and $6\times10^{-6}$, respectively, for a
standard gas-to-dust ratio of 100 by mass. 
The horizontal shaded rectangle indicates the measured value of
$3.8\times10^4$ $L_{\odot}$, and the vertical shaded rectangle highlights
the observationally favored $\tau_{\rm 448}\gtrsim0.2$, corresponding to
$\tau_{\rm 100}\gtrsim8$.

At low column densities, $L_{\rm{H_2O\,448}}$ increases sharply
with $\tau_{448}$ due to the enhancement of the far-IR radiation field,
responsible for the H$_2$O excitation, and to the increase of
$N_{\mathrm{H_2O}}$. The H$_2$O\,448\,GHz line is not masing, but 
usually shows suprathermal excitation ($T_{\mathrm{EX}}>T_{\mathrm{dust}}$) in
some shells. 

The excitation is dominated in all cases by
radiative pumping through the $4_{23}-3_{12}$ and $4_{23}-4_{14}$ lines at
$78.7$ and $132.4$\,$\mu$m (Figure~\ref{fig:levels}). 
Collisional excitation (included in the
models with $n_{\mathrm{H2}}=3\times10^4$\,cm$^{-3}$ and
$T_{\mathrm{gas}}=150$\,K) has the effect of increasing the population
of the low-lying levels from which the radiative pumping cycle works (see
\citetalias{GonzalezAlfonso2014}) thus still having an overall effect on line
fluxes. As $\tau_{100}$ increases above unity, the increase in $\tau_{100}$ does
hardly enhance the far-IR radiation field and 
$L_{\rm{H_2O\,448}}$ flattens. It is just in this regime
where $L_{\rm{H_2O\,448}}$ approaches the observed value 
for high enough $T_{\mathrm{dust}}\gtrsim65$ K or 
$N_{\mathrm{H_2O}}=8\times10^{18}\times\tau_{100}$,
indicating that 
{\it the H$_2$O 448\,GHz line is an excellent probe of buried galaxy nuclei}. 
At higher $\tau_{100}$, line opacity effects
and extinction effects at 448\,GHz (for $\tau_{448}$ approaching unity)
decrease $L_{\rm{H_2O\,448}}$.

With an adopted H$_2$O abundance of $1.5\times10^{-6}$ and
$T_{\mathrm{dust}}\sim65$ K 
(green lines and symbols), we can approximately match the observed 
H$_2$O\,448\,GHz emission (Figure~\protect\subref{fig:model_a}), and the 
448 and 244\,GHz continuum emission (Figure~\protect\subref{fig:model_b})
for $\tau_{448}\approx0.3$ and the observed size.
However, the same observables can also be fitted, for $\tau_{448}=0.4-0.6$,
with a higher $X_{\mathrm{H_2O}}=6\times10^{-6}$  and a more moderate
$T_{\mathrm{dust}}=50$ K (red-dashed lines). 
We can discriminate between both solutions by noting that the dust opacity
conditions required for the H$_2$O\,448\,GHz line to 
emit efficiently, $\tau_{100}>1$, are similar to the conditions required to
have strong absorption in the high-lying H$_2$O lines at far-IR wavelengths
(e.g., \citealt{GonzalezAlfonso2012,Falstad2017}), 
{\it strongly suggesting that both the 448 GHz emission line and the far-IR
  absorption lines arise in similar regions}. 
One of the main H$_2$O lines responsible for the pumping of the 
H$_2$O\,448\,GHz transition, the $4_{23}-3_{12}$ line at $\approx79$\,$\mu$m
(Figure~\ref{fig:levels}), was observed with 
{\it Herschel}/PACS \citep{Pilbratt2010Herschel,Poglitsch2010PACS} within the
open time program HerMoLIRG (PI: E. Gonz\'alez-Alfonso; OBSID=1342248549). 
We compare in Figure~\protect\subref{fig:model_c} the predicted absorbing flux
in this line and the observed value ($-920$ Jy\,km\,s$^{-1}$ between $-150$
and $+150$ km\,s$^{-1}$, the observed velocity range of the H$_2$O\,448\,GHz line at
zero intensity; see Figure~\ref{fig:h2o_profile}). While the
$T_{\mathrm{dust}}\sim50$ K model 
underpredicts the pumping H$_2$O 79\,$\mu$m absorption, the
$T_{\mathrm{dust}}\sim65$ K model better accounts for it, with still some
unmatched redshifted absorption (see insert in
Figure~\protect\subref{fig:model_c}). 
We thus conclude that {\it the H$_2$O 448\,GHz line originates in warm regions
  ($T_{\mathrm{dust}}\gtrsim60$ K)}.

Our favored models indicate that the luminosity of the nuclear region
where the H$_2$O 448\,GHz arises is $(4-6)\times10^{10}$\,\Lsun, i.e.
$\sim25$\% of the total galaxy luminosity. While approximately accounting for
the observables reported in this {\it Letter} 
($L_{\rm{H_2O\,448}}$, allowed $\tau_{448}$ range, $f_{448}$, $f_{244}$, and 
$4_{23}-3_{12}$ absorption strength for the observed size), we advance the
{\it Herschel} detection of very-high lying H$_2$O absorption lines 
indicating the presence of an additional warmer component in
the nuclear region of ESO~320-G030. The full set of H$_2$O (and OH) lines will
be studied in a future work.

\section{Conclusions}

We detected the ortho-H$_2$O \waterline\ transition at 448\,GHz using ALMA
observations of the local spiral LIRG ESO~320-G030. The H$_2$O 448\,GHz
emission arises from the highly obscured nucleus of this galaxy and is
spatially resolved ($r\sim30$\,pc). The H$_2$O 448\,GHz velocity field is
compatible with the global regular rotation pattern of the molecular and
ionized gas in ESO~320-G030. Our radiative transfer modeling shows that it is
mainly excited by the intense far-IR radiation field present in the nucleus of
this source. The conditions for the excitation of the 448\,GHz water
transition indicate that it can probe deeply buried, warm environments both
locally and at high redshifts.

\begin{acknowledgements}

We thank the anonymous referee for useful comments and suggestions.
We thank M.\,Villar-Mart\'in and S.\,Motta for useful comments and careful reading of the manuscript. MPS acknowledges support from STFC through grant ST/N000919/1, the John Fell Oxford University Press (OUP) Research Fund and the University of Oxford. 
EGS, AU, SGB, JMP, LC, AAH, SA, SC, and FRV acknowledge financial support by the Spanish MEC under grants ESP2015-65597-C4-1-R, AYA2012-32295, ESP2015-68694, AYA2013-42227-P and AYA2015-64346-C2-1-P, which is partly funded by the FEDER programme. EGA a Research Associate at the Harvard-Smithsonian CfA and acknowledges support by NASA grant ADAP~NNX15AE56G.
This paper makes use of the following ALMA data: ADS/JAO.ALMA\#2016.1.00263.S. ALMA is a partnership of ESO (representing its member states), NSF (USA) and NINS (Japan), together with NRC (Canada) and NSC and ASIAA (Taiwan) and KASI (Republic of Korea), in cooperation with the Republic of Chile. The Joint ALMA Observatory is operated by ESO, AUI/NRAO and NAOJ.
\end{acknowledgements}

\end{document}